# DATA ACQUISITION SYSTEM FOR THE TUNKA-133 ARRAY

N.M. BUDNEV[2], O.B. CHVALAEV[2], O.A. GRESS[2], N.N. KALMYKOV[1],
V.A. KOZHIN[1], E.E. KOROSTELEVA[1], L.A. KUZMICHEV[1],
B.K. LUBSANDORZHIEV[3], R.R. MIRGAZOV[2], G. NAVARRA[5], M.I. PANASYUK[1],
L.V. PANKOV[2], V.V. PROSIN[1], V.S. PTUSKIN[4], Y.A. SEMENEY[2],
A.V. SKURIKHIN[1], B.A. SHAIBONOV(JUNIOR)[3], CH. SPIERING[6],
R. WISCHNEWSKI[6], I.V. YASHIN[1], A.V. ZABLOTSKY[1], A.V.ZAGORODNIKOV[2]

*1 Scobeltsyn Institute of Nuclear Physics of Moscow State University, Moscow, Russia*
*2 Institute of Applied Physics of Irkutsk State University, Irkutsk, Russia*
*3 Institute for Nuclear Research of Russian Academy of Science, Moscow, Russia*
*4 Institute of Terrestrial Magnetism, Ionosphere and Radiowave Propagation of*
*Russian Academy of Science (IZMIRAN), Troitsk, Moscow Region, Russia*
*5 Dipartimento di Fisica Generale Universita di Torino and INFN, Torino, Italy*
*6 Deutsches Elektronen Synchrotron (DESY), Zeuthen, Germany*

The new EAS Cherenkov array Tunka-133, with about 1 km$^2$ sensitive area, is being installed in the Tunka Valley. The investigated energy range is $10^{15} - 10^{18}$ eV. It will consist of 133 optical detectors based on EMI-9350 PMT. Optical detectors are grouped into 19 clusters with 7 detectors each. The detectors are connected to the cluster box with RG-58 cables. Every PMT signal is digitized in the cluster box with 200 MHz FADC. The cluster boxes are connected to the data acquisition center with a 1 Gb/s optical link. A detailed description of the data acquisition system (DAQ) is presented.

1. **Introduction**

The elaborate study of the energy range $10^{16} - 10^{18}$ eV is of crucial importance for understanding of the origin and propagation of cosmic rays in the Galaxy. The maximum energy of cosmic rays accelerated in SN remnants seems to be in this energy range [1]. As pointed out in [2], in this energy range the transition from Galactic to extragalactic cosmic rays may occur. The new EAS Cherenkov array under construction in the Tunka Valley (50 km from Lake Baikal), with 1 km$^2$ area, was named Tunka-133 [3, 4]. It will allow studying cosmic rays by covering with a single method uniformly the energy range from $10^{15}$ to $10^{18}$ eV. Tunka Valley is known for its good weather conditions (especially during winter). Various EAS Cherenkov arrays – from Tunka-4 [5] to Tunka-25 [6,7] – have been operated at this place.

## 2. The Tunka-133 Array

The Tunka-133 array will consist of 133 optical detectors based on of PMT EMI 9350 with hemispherical photocathode of 20 cm diameter. The 133 detectors are grouped in 19 clusters, each composed of seven detectors – six hexagonally arranged detectors and one in the center. The distance between the detectors is 85 m, similar to Tunka-25. Due to this fact the accuracy of EAS parameter reconstruction will be the same as for Tunka-25, which however have only a sensitive area of 0.1 km$^2$. The accuracies of the core location and shower energy determination are ~6m and ~15% respectively. The accuracy of the EAS maximum depth $X_{max}$ determination (from the Cherenkov light lateral distribution steepness and pulse duration) is ~25 g/cm$^2$. The energy threshold of the array is close to $10^{15}$ eV. During one year operation (400 hours) Tunka-133 will record ~5·10$^5$ events with energies above 3·10$^{15}$ eV, ~300 events with energies higher than $10^{17}$ eV and a handful events with energies higher than $10^{18}$ eV. In addition to the Cherenkov detectors, 5-7 Auger-like water tanks (S = 10 m$^2$, depth = 90 cm) will be constructed for common operation with the Cherenkov array.

## 3. DAQ system

PMT's signals from the optical detectors are fed to the cluster DAQ where they are digitized by custom-made 4-channel-FADC VME modules with 200 MHz sampling rate, and connected to comparators with adjustable thresholds. A trigger system is implemented in each cluster; the cluster trigger is formed by a coincidence of pulses above threshold within a time window of 0.5 μs from at least three optical detectors. The arrival time of the local trigger signal is measured by a cluster timer precisely synchronized with the central DAQ.

Each cluster is connected to the central DAQ by a multi-wire cable with four copper wires (for power supply voltages) and four optical fibers. Optical transceivers operating at 1 GHz are responsible for data transmission and synchronization of the cluster's clock. In the acquisition center optical links are connected to DAQ boards which are strictly synchronized between themselves. A DAQ board provides an interface between the 4 optical lines and 100 Mbit/s Ethernet within the cluster DAQ and is connected to the master PC by a switch.

### 3.1. *Optical detector*

An optical detector (fig. 1) consists of a metallic cylinder of 50 cm diameter, containing the PMT. The container has a Plexiglas window on the top, heated against frost. The angular aperture is defined by geometric shadowing of the PMT. The efficiency is close to 100% up to 30° and reduces to 50% at zenith angles > 45°. The detector is equipped with a remotely controlled lid protecting the PMT from sunlight and precipitation. Apart from the PMT with its high voltage supply and preamplifiers, the detector box contains a light emitting diode (LED) for both amplitude and time calibration and a controller. The controller is connected with the cluster electronics by twisted pair (RS-485). To provide the necessary dynamic range of $10^4$, two analog signals, one from the anode and another one from the dynode, are read out. They are amplified and then transmitted to the central electronics hut of each cluster. The amplitude ratio of these signals is about 30. It is not planned to heat the inner volume of the optical detector boxes, therefore all the detector electronics is designed to operate over a wide temperature range (down to –40°C).

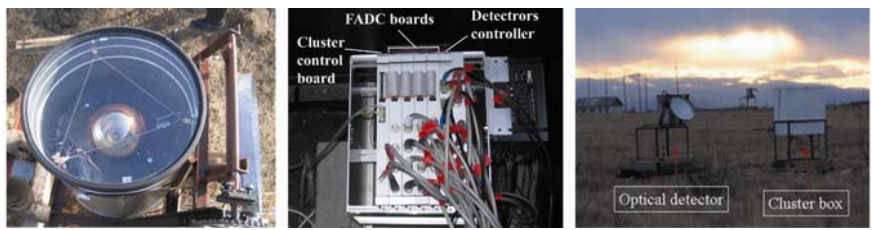

Fig. 1. Optical detector and cluster box photos (from left to right: top view of the optical detector, the cluster DAQ , the cluster center).

### 3.2. *The cluster DAQ system*

The cluster DAQ (fig. 1) consists of 4 FADC boards, a detector control board and a cluster control board. The detector control board provides extra service by controlling the detector and synchronously firing the calibration LEDs. Optical transceivers, trigger module, time synchronization module and local timer are included inside the cluster control board, based on a FPGA (XLINX Spartan XC3S200).

Every FADC board consists of four simultaneously digitizing ADCs which are logically divided into two channels (one per the anode and dynode signals from each optical detector), see fig. 2. Each FADC has 12 bit resolution and samples at 200 MHz. The digitized signal from each ADC is transferred to the FPGA which handles the data. A double-buffered FIFO memory for 1024 counts was implemented inside the FPGA: while one buffer is waiting for readout, the second one is connected to the ADC outputs, thus minimizing the readout dead-time. After arrival of the trigger signal from the master device, 512 counts before and 512 after the trigger are readout. Only one signal (anode) from each channel is connected to the comparator with digitally adjustable threshold, which produces the request signal to the trigger module. The anode signal is also connected to an amplitude analyzer, accumulating monitor histograms for every channel.

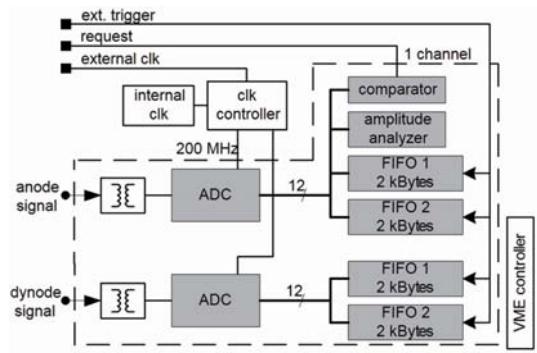 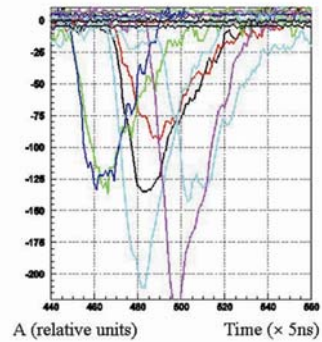

Fig. 2. Internal structure of FADC board (shown only one channel; the second one is identical).

Fig. 3. Example of an event in a cluster (energy $(5\pm2)\cdot10^{17}$ eV)

### 3.3. *Time synchronization system*

Optical detectors (anode and dynode signals) are linked to the cluster DAQ with RG58 cables which have identical length, so they will not be further considered here. But the clusters are connected to the central DAQ via optical cables with unknown lengths. Every cluster has a local timer, used for the cluster event time; therefore offsets between cluster timers must be calibrated.

The time synchronization is implemented in two stages (fig. 4). Cable length and other propagation delays are measured in the first stage. The center sends to the cluster a special packet, returned by the cluster after decoding. The total time delay, divided by two, corresponds to the propagation delay. On the second stage, the center sends a timer reset command simultaneously to all cluster DAQs. The knowledge of the propagation delay allows to calculate differences between timers for each cluster. The accuracy of this time synchronization is about 10 ns. To avoid any mis-timing, the synchronization is being checked once per second by hardware.

55

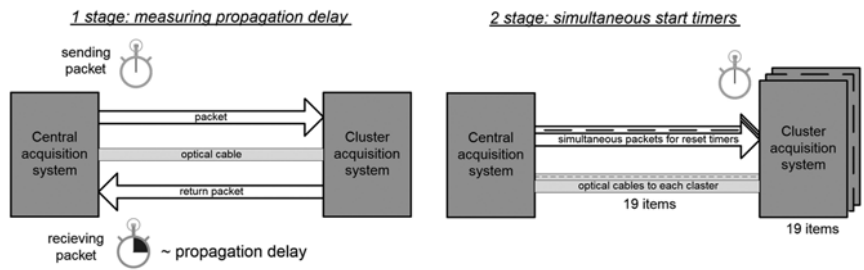

Fig. 4. Concept of the time synchronization system.

## 4. Conclusion

A 1-km$^2$ Cherenkov EAS array is under construction in the Siberian Tunka Valley. A data acquisition system was built and is under comprehensive testing now. The first cluster of the TUNKA-133 array (seven optical detectors) has been successfully operated during the winter 2006 - 2007.

During the winter season 2007-08, four clusters will operate. Commissioning of the main part of the array is planned in autumn of 2008.

**Acknowledgments**

The present work is supported by the Russian Ministry of Education and Science, by the Russian Fund of Basic Research (grants 05-02-04010, 05-02-16136, 06-0216520, 07-02-00904) and by the Deutsche Forschungsgemeinschaft DFG (436 RUS 113/827/0-1).